\documentclass[doublecol]{epl2}

\begin{document}

\title{Thermoelectricity of Wigner crystal in a periodic potential}

\author{O.V.Zhirov\inst{1,2}\and  D.L.Shepelyansky\inst{3}} 
\shortauthor{O.V.Zhirov, D.L.Shepelyansky}
\institute{
 \inst{1} Budker Institute of Nuclear Physics, 630090 Novosibirsk, Russia\\
 \inst{2} Novosibirsk State University, 630090 Novosibirsk, Russia\\
 \inst{3} Laboratoire de Physique Th\'eorique du CNRS (IRSAMC), 
  Universit\'e de Toulouse, UPS, F-31062 Toulouse, France
}

\abstract{We study numerically the thermoelectricity of
the classical Wigner crystal placed in  a periodic potential and
being in contact with a thermal bath modeled by the Langevin dynamics.
At low temperatures the system has sliding and pinned
phases with the Aubry transition between them.
We show that in the Aubry pinned phase
the dimensionless Seebeck coefficient can reach  very high 
values of several hundreds.
At the same time the charge and thermal conductivity 
of crystal drop significantly inside this phase. 
Still we find that the largest values of  $ZT$ factor 
are reached in the Aubry phase 
and for the studied parameter range
we obtain $ZT \leq 4.5$.  We argue that this system
can provide an optimal regime for reaching high
$ZT$ factors and  realistic modeling  
of thermoelecriticy. 
Possible experimental realizations of this
model are discussed.
}
\pacs{84.60.Rb}{Thermoelectric, electrogasdynamic and other direct energy conversion}
\pacs{72.20.Pa}{Thermoelectric and thermomagnetic effects}
\pacs{73.20.-r}{Electron states at surfaces and interfaces}


\maketitle

\section{Introduction}

Computer microelectronic elements go to nanoscale sizes and control
of electrical currents and related heat flows becomes a technological challenge 
(see e.g. \cite{majumdar2004,thermobook}). 
By the thermoelectric effect a temperature difference $\Delta T$
generates an electrical current that can be compensated by a voltage difference
$\Delta V$. The ratio $S=\Delta V/ \Delta T$ is known as the Seebeck coefficient,
or thermopower, which plays an important role 
in the thermoelectric material properties. 
The thermoelectric materials are ranked by a figure of merit factor 
$ZT=S^2\sigma T/ \kappa$ \cite{ioffe}, 
where $\sigma$ is the electric conductivity,
$T$ is material temperature and $\kappa$ is the thermal conductivity.
To be competitive with usual refrigerators one needs to find materials
with $ZT > 3$ \cite{majumdar2004}. 
Various experimental groups try to reach this high value
by skillful methods trying to reduce the  thermal conductivity $\kappa$
of samples keeping high electron conductivity $\sigma$ and high $S$
(see e.g. 
\cite{ztexp2001,ztexp2008nanowire1},
\cite{ztexp2008nanowire2,ztexp2008bism},\cite{ztexp2012}).
At room temperature 
the maximal values $ZT \approx 2.4$ have been reached 
in semiconductor superlattices
\cite{ztexp2001} while for silicon nanowires 
a factor $ZT \approx 1$  has been demonstrated 
\cite{ztexp2008nanowire1,ztexp2008nanowire2}.
This shows that the volume reduction allows to
decrease the thermal conductivity of lattice phonons
and increase $ZT$ values. 

It is interesting to consider the situations
when the contribution of lattice phonons
is completely suppressed to see if in such
a case one can obtain even larger $ZT$ factors.
Such extreme regime can be realized with 
an electron gas, e.g. in two dimensions (2DEG),
where at $T \sim 1 K$ a contribution of 
lattice phonons is completely suppressed.
In such a regime recent experiments \cite{pepper2012}
reported giant Seebeck coefficients
$S \sim 30 mV/K$ obtained in a high resistivity domain.

While it is  challenging to eliminate
the contribution of lattice phonons experimentally
it is rather easy to realize such a situation in numerical
simulations simply replacing a lattice of atoms by a 
fixed periodic potential. After that we are faced the problem
of thermoelectricity of Wigner crystal
in a periodic potential. In this Letter we study this
problem in one dimension (1D), which can be viewed
as a mathematical model of silicon nanowires. We note that
the ground state and low temperature properties of
this system in classical and quantum 
regimes have been investigated in \cite{fki}.
It has been shown that at a typical incommensurate
electron density the Wigner crystal slides easily 
in a potential of weak amplitude
while above a critical amplitude  
the electrons are pinned by a lattice.
The results \cite{fki} show that the properties
of the Wigner crystal are similar to those of the
Frenkel-Kontorova model where the transition
between sliding and pinned phases is known as
the Aubry transition \cite{aubry}
(see detailed description in \cite{braun}).
The positions of electrons 
on a periodic lattice are locally described by 
the Chirikov standard map \cite{chirikov,scholar}.
Similar dynamical properties appear
also for the Wigner crystal in wiggling snaked
nanochannels \cite{snake}. 

The previous studies of the Wigner crystal in a
periodic potential \cite{fki}
have been concentrated on analysis of the ground state
properties at lower temperatures.
Here we analyze the transport properties of the
crystal at finite temperatures studying its 
electron and thermal conductivities.
Our approach allows to obtain the Seebeck
coefficient and the figure of merit $ZT$
at different regimes and various parameters.
We note that there has been a significant interest to
the heat transport and thermal conductivity 
in nonlinear lattices \cite{politi,baowenrmp}
but till present there have been no studies of
thermoelectricity of interacting electrons
in  periodic lattices. We present the 
investigations of this generic case in this Letter.

\section{Model description}
The Hamiltonian of the 1D Wigner crystal in a periodic potential reads:
\begin{eqnarray}
\label{eq1}
 H&=&\sum_i \left(\frac{p^2_i}{2} + K\cos x_i
   +\frac{1}{2}\sum_{j\neq i} \frac{1}{|x_i-x_j|}\right) ,
\end{eqnarray} 
where $x_i,p_i$ are coordinate and momentum of electron $i$,
$K$ is an amplitude of periodic potential or lattice.
As in \cite{fki} we use the units with
$e=m=k_B=1$, where $e$ and $m$ are electron charge
and mass, $k_B$ is the Boltzmann constant, the lattice period is $2\pi$.
The rescaling back to physical units is given in \cite{fki}.
It is interesting to note that at $e=k_B=1$ we have $S$ as a dimensionless
coefficient, e.g. $S=30 mV/K$ from \cite{pepper2012}
corresponds to $S = 2585$. Generally, in an ergodic regime 
induced by a developed dynamical chaos or thermal bath, 
one expects to have
$S \sim 1$ since a variation of potential or temperature
should produce approximately  the same charge redistribution.
Thus, in our opinion, large values of dimensionless
Seebeck coefficient $S$ indicate a strongly nonergodic regime
of system dynamics. We will see below confirmations of this 
statement.

We concentrate our studies on a case of typical
irrational electron density $n_e=\nu/2\pi$, per lattice period, 
given by the golden rotation number $\nu=\nu_g=1.618...$.
As in \cite{fki} we use the Fibonacci rational
approximates with $N$ electrons $(0 \leq i \leq N-1)$ 
on $M$ lattice periods (e.g. $34$ and $21$ or $55$ and $34$).

According to \cite{fki} the Aubry transition at density $\nu_g$ takes place
at $K=K_c=0.0462$ so that the Wigner crystal is in a sliding phase for
$K<K_c$ and it is pinned by the potential at $K>K_c$. In the latter
case there are exponentially many static configurations
being exponentially close in energy  to the Aubry cantori ground state.
The sliding phase corresponds to the continuous 
Kolmogorov-Arnold-Moser (KAM) curves with $\nu_g$
rotation number.

\begin{figure}[h]
\begin{center}
\includegraphics*[width=7.8cm]{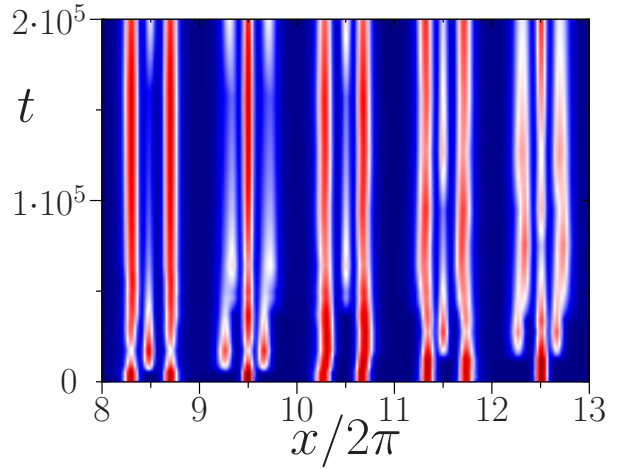}
\end{center}
\vglue -0.3cm
\caption{Electron density variation 
in space and time from one Langevin trajectory at
$K/K_c=2.6$, $T/K_c=0.11$, $\eta=0.02$,
$N=34$, $M=L/2\pi=21$; density changes
from zero (dark blue) to maximal density (dark red);
only a fragment of $x$ space is shown.}  
\label{fig1}
\end{figure}

To study the thermoelectic effect (\ref{eq1})
we add interactions with a substrate, which plays a role of a thermal
bath with a given temperature distribution $T(x)$ along $x$-axis of the
electron chain. We  also add a static electric field $E_{dc}$. The thermal
bath is modeled by the Langevin force (see e.g. \cite{politi})
so that the equations 
of electron motion are:
\begin{equation}
\dot{p}_i = -\partial H/\partial x_i +E_{dc} -\eta p_i+g \xi_i(t)
\; , \;\; \dot{x_i} = p_i \; .
\end{equation}
Here, the parameter $\eta$ phenomenologically describes 
dissipative relaxation
processes, and the amplitude of Langevin force is given 
by the fluctuation-dissipation theorem $g=\sqrt{2\eta T}$.
The normally distributed random variables $\xi_i$ are 
as usually defined by correlators
$\langle\xi_i(t)\rangle=0$,
$\langle\xi_i(t) \xi_j(t')\rangle=\delta_{i j}\delta(t-t')$.
The  time evolution is obtained by the 4th order  Runge-Kutta integration
with a time step $\Delta t$, at each such a step
the Langevin contribution is taken into account. We checked that
the results are not sensitive to the step $\Delta t$ by its 
variation by a factor ten, the data
are mainly obtained with $\Delta t =0.02$. 
We use the hard wall boundary conditions for
electrons at the ends of the chain  $x=0; L$ with the total system length
$L=2\pi M$. We also note that the Coulomb interaction 
couples all electrons in the sample. However, the results
of \cite{fki,snake} show that only nearest neighbors 
are effectively count. Due to that we present the numerical results
for this approximation. We ensured that our results are not 
sensitive to including other neighbors.

A typical variation of electron density in space $x$ and time $t$
is shown in Fig.~\ref{fig1} for the Aubry pinned phase.
Transitions,  induced by thermal fluctuations,
from one to two electrons inside one potential minimum
are well visible.

\section{Numerical results for Seebeck coefficient}

\begin{figure}[h]
\begin{center}
\includegraphics*[width=4cm]{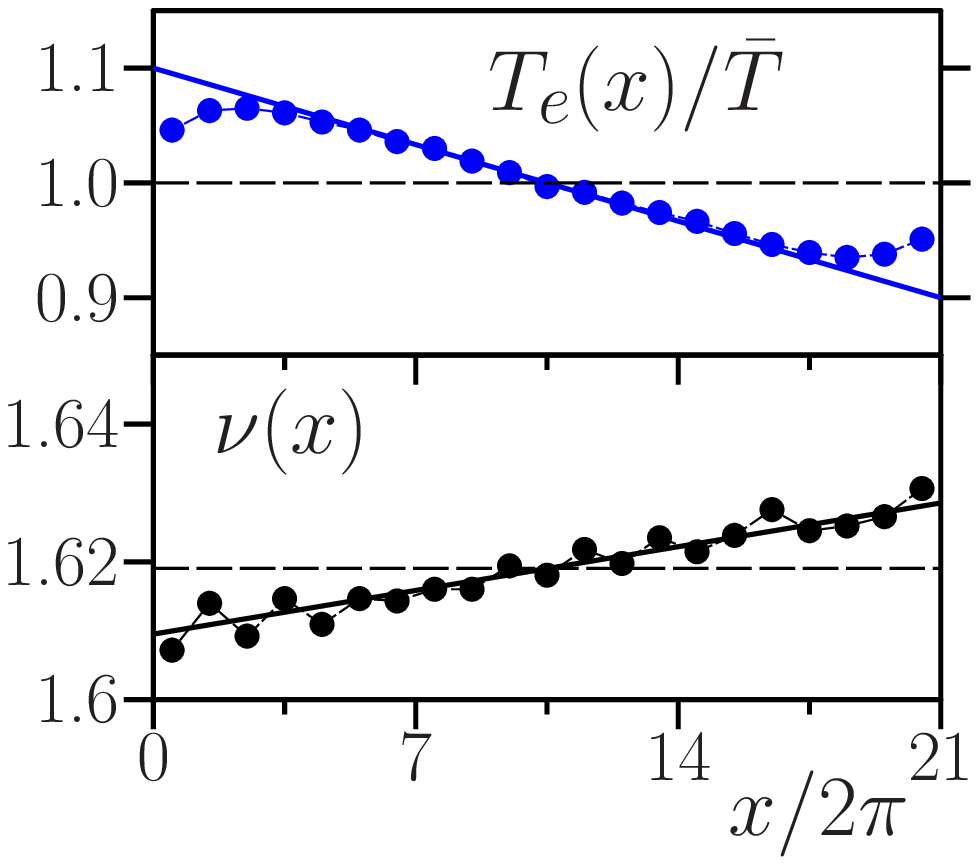}
\includegraphics*[width=4cm]{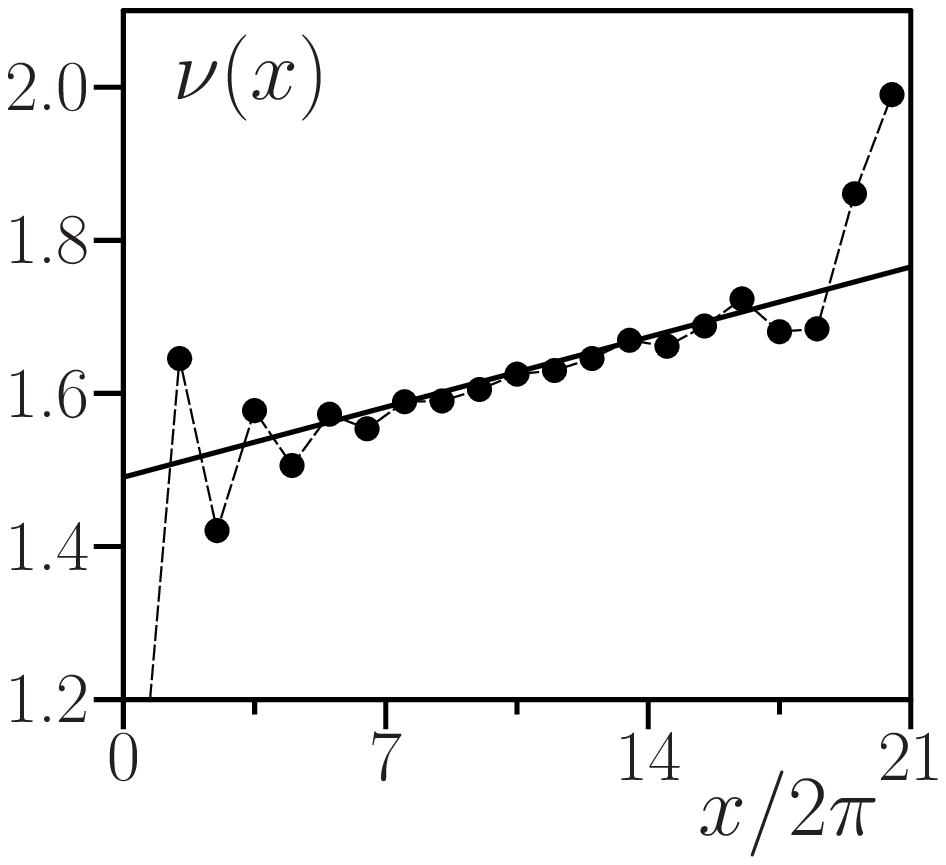}
\end{center}
\vglue -0.3cm
\caption{{\it Left panels}: dependence of electron temperature
$T_e(x)$ {\it (top, blue points)} and rescaled density $\nu(x)$ 
{\it (bottom, black points)}
on distance $x$ along the chain placed on the Langevin substrate with
a constant temperature gradient (it is shown by the blue line)
at average temperature $\bar{T}=0.01$
and temperature difference $\Delta T = 0.2  \bar{T}$;
black line shows the fit of density variation in the bulk part of the sample.
{\it Right panel}: density variation produced by
a static electric field $E_{dc}=4 \times 10^{-4}$
at a constant substrate temperature $T=0.01$;
black line shows the fit of gradient in the bulk part of the sample.
Here $N=34, M=21$, $K=1.52K_c$, $\eta=0.02$, averaging 
is done over time interval
$t=10^7$; $S=3.3$ at $T=0.01 \approx 0.22 K_c$}  
\label{fig2}
\end{figure}

To compute $S$ we impose a constant temperature gradient
on the Langevin substrate with a temperature difference $\Delta T$
at the sample ends. Then we compute the local
electron temperature $T_e(x) = \langle p^2(x)\rangle_t$
where the time average of electron velocities are
done over a large time interval with up to $t=10^7$.
To eliminate periodic oscillations along the chain we
divide it on $M$ bins of size $2\pi$
and do all averaging inside each bin. Typical examples of
variations of electron temperature $T_e(x)$ and electron rescaled density
$\nu(x)=2\pi n_e(x)$  along the chain are shown
for a given $\Delta T$  in Fig.~\ref{fig2} (left panels).
The chain ends are influenced by the boundary conditions,
but in the main bulk part of the sample we obtain a linear
gradient variation of $T_e(x)$ and $\nu(x)$. The linear fit 
of $T_e(x)$ and $\nu(x)$ in the bulk part allows to determine 
the response of the Wigner crystal on substrate temperature variation.
In a similar way at fixed substrate temperature $T$ we 
can find the  density variation $\nu(x)$ induced by a static field $E_{dc}$
at the voltage difference $\Delta V= E_{dc} L$,
as it is shown in Fig.~\ref{fig2} (right panel).
For the computation of $S$ we find convenient to
apply such a voltage $\Delta V$ which at fixed $T$
creates  the same
density gradient as those induced by temperature difference $\Delta T$
at $E_{dc}=0$. Then by definition $S=\Delta V/\Delta T$.
The data are obtained in the linear response regime
when $\Delta T, E_{dc}$ are sufficiently small.

\begin{figure}[h]
\begin{center}
\includegraphics*[width=4cm]{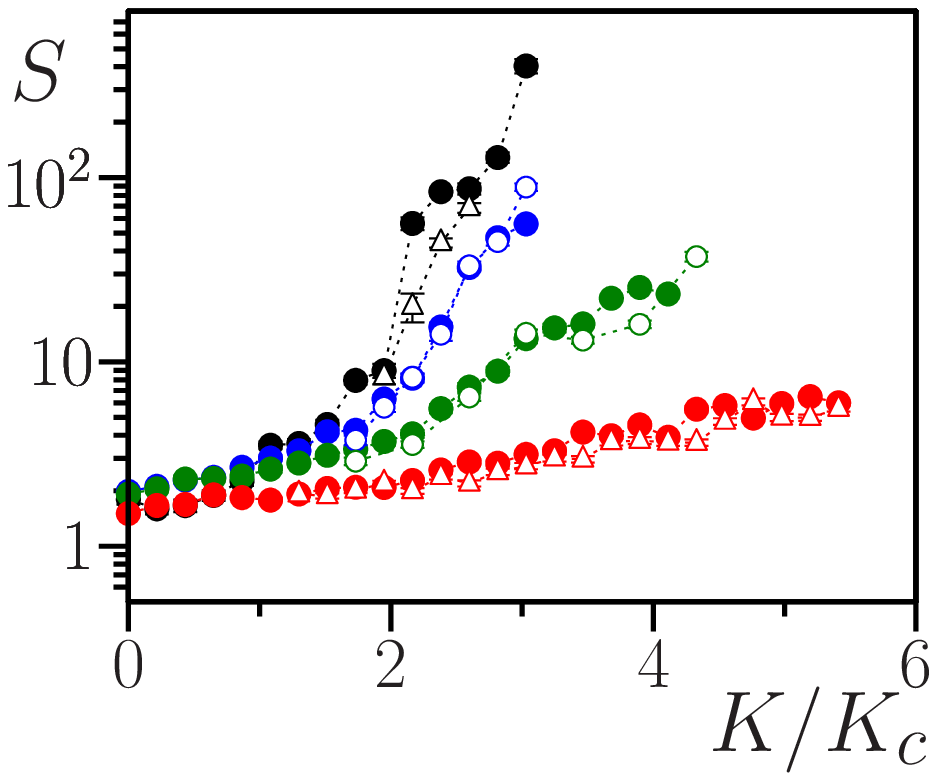}
\includegraphics*[width=4cm]{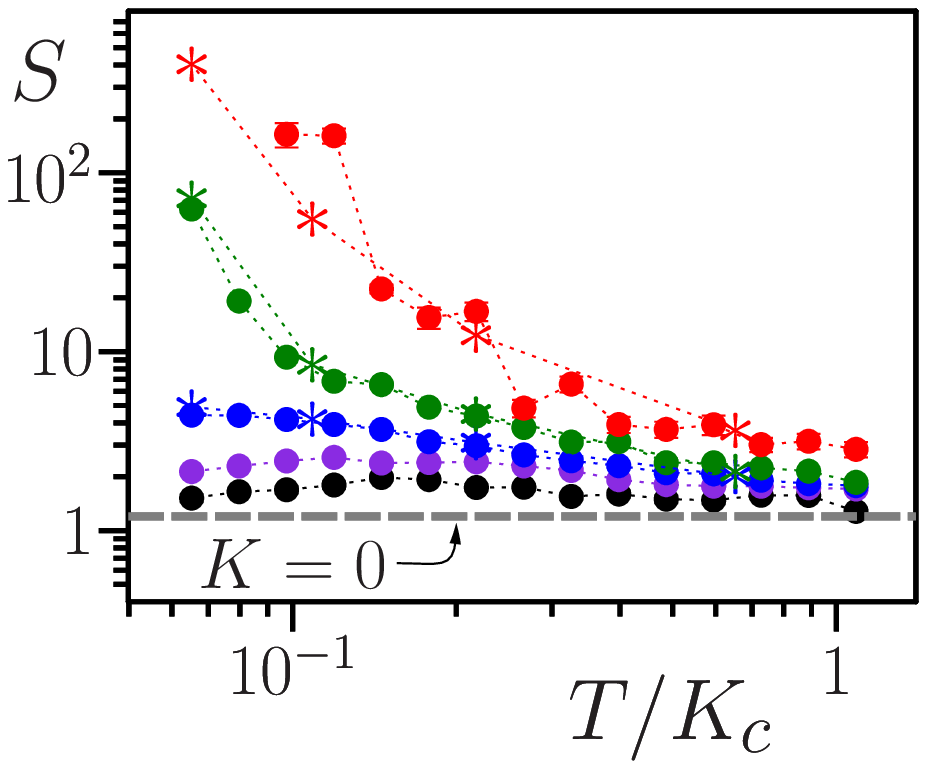}
\end{center}
\vglue -0.3cm
\caption{{\it Left panel}: Dependence of the Seebeck coefficient 
$S$ on rescaled potential amplitude $K/K_c$ at
temperatures $T/K_c=0.065, 0.11, 0.22$ and $0.65$ shown by
black, blue, green and red colors, respectively from top to bottom.  
The full and open symbols correspond respectively to chains 
with  $N=34$, $M=21$ and $N=55$,  $M=34$.
{\it Right panel}: Dependence of $S$ on $T/K_c$ at different  
$K/K_c=0, 0.75, 1.5, 2.2, 3$ shown respectively by black,  violet, blue, green
and red points; $N=34$, $M=21$; the dashed gray line 
shows the  case $K=0$ for noninteracting particles.
The stars show corresponding results from
left plane at same $N, M$. Dotted curves are drown to adapt an eye.
Here and in other Figs. the statistical error bars are shown when 
they are larger than the symbol size. Here $\eta=0.02$.
\label{fig3}}
\end{figure}

The dependencies of obtained Seebeck coefficient $S$ on
$K$ and $T$ are presented in Fig.~\ref{fig3}. The data show that
at $K<K_c$ we have $S \sim 1$ practically for all
temperatures. Here the Lanvegin thermostat
efficiently produces an ergodic distribution over all
configurations of electrons and we have $S \sim 1$
in agreement with the above ergodic argument.
For $K>K_c$ we find a significant increase of $S$
at low temperatures $T<K_c$. In this regime the crystal
is pinned by the lattice and different 
configuration states are separated by
potential barriers $\Delta U \sim K-K_c$
so that the transitions between configurations
are suppressed by the Boltzmann factor $\exp(-\Delta U/T)$.
Thus here long times are needed to
have a transition between configurations \cite{fki}.
In such a regime large voltage $\Delta V$
is required to produce the same density gradient
as those given by a fixed $\Delta T$. This leads
to large $S$ values generated by big and rare thermal fluctuations.

To check the stability of obtained results in the nonergodic regime with large
$S$ we use three different numerical methods:

(a) {\it cold start} from the Aubry
ground state at a given $K$ and $T=0$, 
followed by a warm up to required $T$ and then computing of
the responses to a temperature gradient or electric field; 
in this approach the system
evolves during a relaxation time 
$t_{rel}\sim  10^6$ until the density
response is stabilized, then the computations 
of gradients are performed on a time scale
$t_{com}$ determined by the condition of
target statistical accuracy
(typically $t_{com} \sim 10^7$);

(b) {\it zero potential start} from the ground state at $K=0$
and given $T$ followed by a sweep over $K$ with a
step $\Delta K$ (typically $\Delta K=0.01$);  at each step 
the responses of current state to $E_{dc}$ or $\Delta T$
are determined;  after $t_{rel}=5 \times 10^4$
the gradients are computed on times $t_{com} \ge 10^4$
determined by target accuracy; next step to $K+\Delta K$
starts from the reached steady state at previous $K$ value,
continuing up to required $K_{max}$ value,
that completes one sweep in $K$; 
then we repeat sweeps about $20$ to $200$ times 
to improve statistical accuracy;

(c) {\it hot start} from the Aubry ground state at given $K$ with
a warm  up to $T_{max}=0.05 \approx K_c$, followed by a sweep from
$T=T_{max}$ down to $T=T_{min}=0.003$ with equidistant steps in $\ln T$,
in a way similar to (b) with a similar number of sweeps.

The data in left and right panels of Fig.\ref{fig3}
are obtained by the methods (b) and (c) respectively. 
The stars  in the right panel show 
the corresponding data from the left plane. A good
agreement between  methods (b) and (c) 
confirms the validity of obtained results.
The results from a more time-consuming method (a) 
give a similar agreement with those 
methods (b),(c) of Fig.~\ref{fig3}
(data not shown). 
The comparison of results with $N=34$ and $55$ electrons
shows their independence of the chain length.
However, at $K \gg K_c$ and $T \ll K_c$
very long computations are required to obtain statistically 
reliable results.

The obtained results show that large values
of $S > 100$ can be reached in the pinned phase
$K > K_c$ at low temperatures. 
The growth of $S$ is roughly
proportional to the inverse Boltzmann
factor. This 
nonergodic regime is characterized
by big fluctuations. We think that a similar regime
appeared in 2DEG experiments 
with even larger values $S \sim 10^3$ \cite{pepper2012}.

\section{Properties of charge and thermal conductivities}

The large values of $S$ do not guaranty high values of 
figure of merit factor $ZT$ which depends also
on charge and thermal conductivities $\sigma$, $\kappa$.

\begin{figure}[h]
\begin{center}
\includegraphics*[width=4cm]{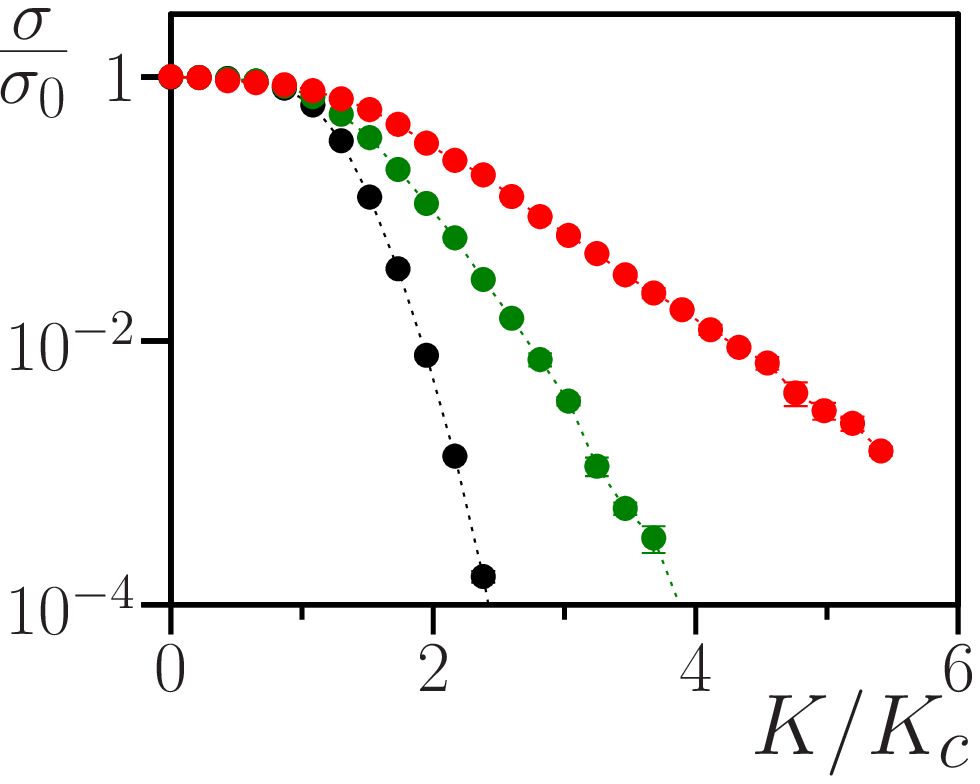}
\includegraphics*[width=4cm]{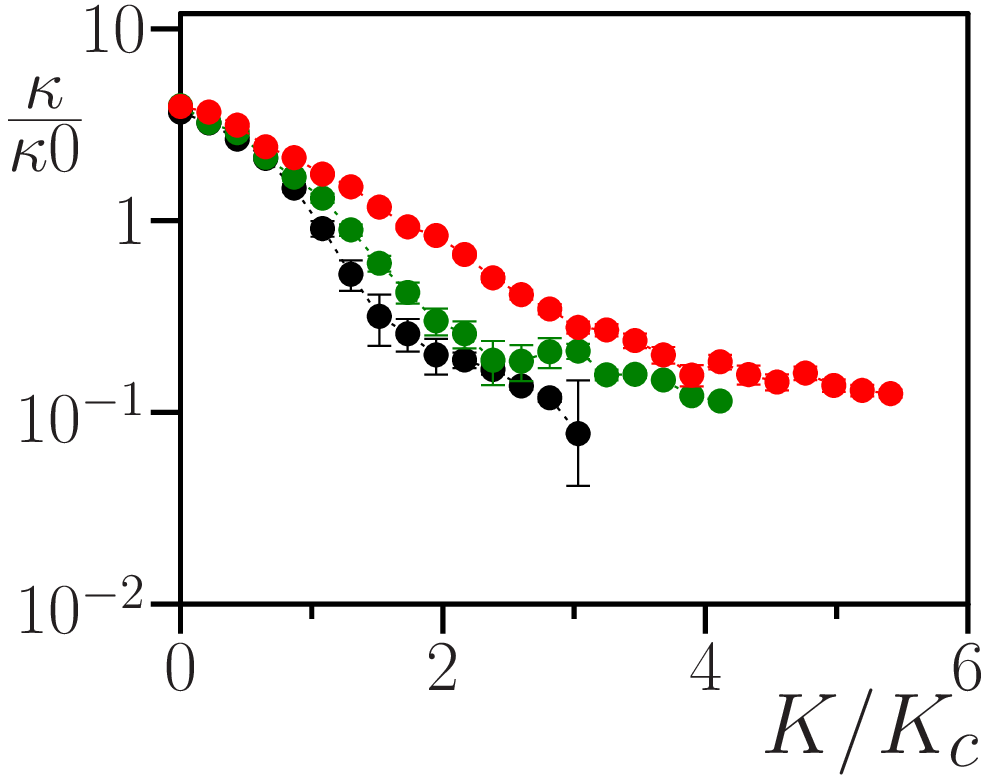}
\end{center}
\vglue -0.3cm
\caption{{\it Left panel}: Rescaled electron conductivity 
$\sigma/\sigma_0$ as a function of $K/K_c$
shown at rescaled temperatures $T/K_c=0.065$, $0.22$, $0.65$
by black, green and red points respectively.
{\it Right panel}: 
Rescaled thermal conductivity $\kappa/\kappa_0$ 
as a function of $K/K_c$ shown at
same temperatures and colors as in left panel.
Here we have $N=34$, $M=21$, $\eta=0.02$,
$\sigma_0=\nu_g/(2\pi \eta)$, $\kappa_0=\sigma_0 K_c$.}
\label{fig4}
\end{figure}

To determine $\sigma$ we use the periodic boundary
conditions  (electrons on a circle) 
and compute the average velocity $v_{el}$ of
the Wigner crystal in a weak electric field $E_{dc}$
(acting along the circle)
being in a linear response regime. The averaging is 
done over a typical time interval $t =10^7$
and over all electrons. Then the charge current is 
$j=n_e v_{el} = \nu v_{el}/2\pi$ and $\sigma=j/E_{dc}$. 
In absence of potential at $K=0$ we have
a crystal moving as a whole with $v_{el}=E_{dc}/\eta$
and corresponding to the conductivity
$\sigma=\sigma_0=\nu_g/ (2\pi \eta)$
($\nu_g \approx 1.618...$). This theoretical result
is well reproduced by numerical simulations as it is
shown in Fig.~\ref{fig4} (left panel).

For $K<K_c$ the conductivity $\sigma$ is practically independent
of $T$, $K$. However, for $K>K_c$ we have a sharp
exponential drop of $\sigma$ with increasing $K$ and decreasing temperature.
This drop is satisfactory described by the thermal activation
dependence $\sigma \propto \exp(-(K-K_c)/T)$,
at least when $K$ is significantly larger $K_c$.  
We note that the
temperature dependence differs significantly from
those in 2DEG experiments \cite{pepper2012}
where resistivity becomes independent of $T$ for $T<1 K$.
We attribute this to  2D features of these experiments
and to quantum effects being important at $T \sim 1 K$.
Indeed, the quantum fluctuations can produce sliding
of the Wigner crystal even in the classically pinned phase
as it is shown for 1D in \cite{fki}.

Another important feature of $\sigma$ variation with the system
parameters is that $\sigma \sim 1/\eta$ for $K<K_c$
and that $\sigma$ is practically independent of $\eta$
for $K>K_c$. There is only a moderate variation of $S^2 \sigma$
by a factor $4$ when $T/K_c$ changes from $0.1$ to $10$.
We discuss this point in more detail later.  

The thermal gradient produces not only the 
charge density variation but also
a heat flow $J$. This flow is related
to the temperature gradient 
by the Fourier law with the thermal
conductivity $\kappa$: $J=\kappa \partial T / \partial x$
(see e.g. \cite{thermobook,politi}). The flow $J$
can be determined from the analysis of
forces acting on a given electron $i$
from left and right sides respectively:
$f_i^{L}=\sum_{j<i} 1/|x_i-x_j|^2$,
$f_i^{R}=-\sum_{j>i} 1/|x_i-x_j|^2$.
The time averaged energy flows, 
from left and right sides,
to an electron $i$ moving with a velocity $v_i$
are respectively
$J_{L,R} = \langle f_i^{L,R} v_i \rangle_t \;$.
In a steady state the mean electron energy is independent of 
time and $J_L + J_R=0$. But the difference of these flows
gives the heat flow along the chain: $J=(J_R-J_L)/2=
 \langle ( f_i^{R}- f_i^{L}) v_i/2 \rangle_t \; $.
This  computation of the heat flow,
done with hard wall boundary conditions,
allows us to determine the thermal conductivity
via the relation $\kappa=J L/\Delta T$. Within
numerical error bars
we find $\kappa$ to be independent of small $\Delta T$
and number of electrons $N$ ($21 \leq N \leq 144$). 

In principle, each electron interacts also with the substrate.
However, in the central part of the chain
the electron temperature is equal to the local
temperature of the substrate due to local thermal equilibrium.
This fact is directly seen in Fig.~\ref{fig2} (left top panel,
{\it cf.} blue points and straight line).
Thus, we perform additional averaging of the heat flow
in the central $1/3$ part of the chain 
improving the statistical accuracy of data.

The dependence of computed thermal conductivity
$\kappa$ on the amplitude of the potential
$K$ is shown in Fig.~\ref{fig4} (right panel).
It is convenient to present 
$\kappa$ via a ratio to $\kappa_0=\sigma_0 K_c$
to have results in dimensionless units.
Similar to the charge conductivity $\sigma$,
we find that $\kappa \approx 3.9 \kappa_0$
at $K < K_c$ being practically independent of 
temperature $T$ for $T < K_c$. 
However, the transition to zero temperature
and $\eta=0$ is singular due to divergence of $\kappa$ in 
weakly nonlinear regular chains 
as discussed in \cite{politi}.

In the pinned phase at $K>K_c$
we see an exponential drop of $\kappa$ with increase of $K$
and decrease of $T$ at $T<K_c$. As for $\sigma$,
we find that for $K>K_c$ the thermal conductivity
is practically independent of dissipation rate $\eta$.
We will discuss this in more detail below.

\section{Results for figure of merit factor $ZT$}

Now we determined all required characteristics
and can analyze what $ZT$ values are typical for
our system and how $ZT$ depends on the parameters.

\begin{figure}[h]
\begin{center}
\includegraphics*[width=8.0cm]{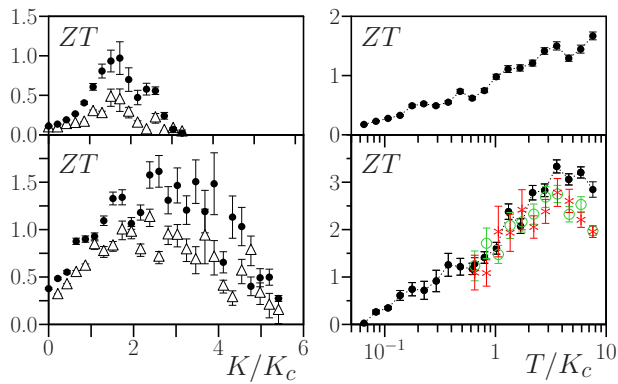}
\end{center}
\vglue -0.3cm
\caption{{\it Left panels}: 
Dependence of $ZT$ on $K/K_c$ at 
temperatures $T/K_c=0.11$ (top panel)  and $T/K_c=0.65$ 
(bottom panel); the black points  and open triangles 
correspond respectively to $\eta=0.02$ and 
$\eta=0.05$ at $N=34$, $M=21$.
{\it Right panels}:  Dependence of $ZT$ on 
$T/K_c$ for  $K/K_c=0.75$  at $\eta=0.02$,
$N=34$, $M=21$.
{\it Bottom right panel}: Same as in
top right panel at $K/K_c=2.6$ and
$N=34$, $M=21$ (black points);
$N=89$, $M=55$ (green circles);
$N=144$, $M=89$ (red stars).} 
\label{fig5}
\end{figure}

The typical  results are presented in Fig.~\ref{fig5}
where at chosen  parameters we have $ZT < 3.5$.
At fixed $T=0.65K_c$ we have an optimal value
of $K$ with a maximum of $ZT$ 
at a certain $K \sim 2K_c$,
its position moves slightly to larger $K$
with an increase of $T$ (left panels).
At fixed $K =2.6 K_c$, taken approximately at the maximum
of $ZT$ (left bottom panel), there is a visible 
logarithmic type growth
of $ZT$ with increasing $T$ approximately 
by a factor $7$ in a range  $0.1 \leq T/K_c \leq 50$
(right panels). 
A further increase up to  $T \gg 5 K_c \approx 0.25 $
is not very interesting since then we 
start to have temperature to be larger than
the energy of Coulomb interaction 
$E_W$ between electrons ($T > E_W =\nu_g/2\pi \approx 0.25 $)
and the model goes to another limit of
rigid type balls which is not very realistic.

\begin{figure}[h]
\begin{center}
\includegraphics*[width=8.0cm]{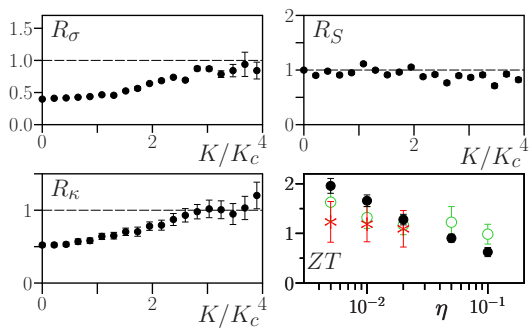}
\end{center}
\vglue -0.3cm
\caption{{\it Left panels}: 
Dependence of ratios $R_\sigma$ (top)
and $R_\kappa$ (bottom) 
on $K/K_c$ at $T/K_c=0.65$.
{\it Right top panel}: Same as in left panels
for ratio $R_S$. All ratios are defined in the text.
{\it Right bottom panel}: dependence
of $ZT$ on   $\eta$ at $T/K_c=0.65$
at $N=34$, $M=21$ (black points); $N=89$, $M=55$
(green circles);  $N=144$, $M=89$ (red stars)
at fixed $K/K_c=2.6$ and $T/K_c=0.65$.} 
\label{fig6}
\end{figure}

The results for two values of dissipation
$\eta = 0.02; 0.05$ shown in Fig.~\ref{fig5}
indicate that $ZT$  drops with increase of $\eta$.
To understand the effects of $\eta$ in a better
way we show the dependence of ratio
$R_S=S(\eta=0.05)/S(\eta=0.02)$
on $K/K_c$ at fixed $T/K_c=0.65$ in Fig.~\ref{fig6}.
The dependence of similar ratios $R_\sigma$ and $R_\kappa$
for $\sigma$ and $\kappa$ are also shown there.
We find   $R_\sigma \approx R_\kappa \approx 0.5$
at $K \ll K_c$ and $R_\sigma \approx R_\kappa \approx 1$
for $K>K_c$. At $K \ll K_c$ the ratios are close to 
the expected value $0.4$ following from the theoretical
scaling $\sigma_0  \propto 1/\eta$
and similar expected dependence $\kappa_0 \propto 1/\eta$.
However, in the pinned phase the
dependence of $\sigma$ and $\kappa$ on $\eta$ practically disappears.
The physical mechanism of this effect is due to the fact
that the electrons are pinned  by the lattice and 
Wigner crystal phonons
are localized,  and 
hence, their mean free path becomes 
smaller than its value at $K=0$ when it is given by 
the dissipative exchange with the Langevin substrate. 
The ratio $R_S$ is not sensitive to the variation of $K/K_c$
even if $S$ changes strongly with $K$ (see Fig.~\ref{fig3}).
A similar behaviour of ratios
is obtained at lower $T/K_c \approx 0.1$
with somewhat more sharp change between
limit values $0.5$ and $1$ around $K/K_c \approx 2$.
We also checked that the ratios constructed for other values
of $\eta$ (e.g. $\eta=0.01, 0.1$, instead of above $\eta=0.05$)
also saturate at unit value for $K/K_c >2$.
Thus, at $K/K_c>2$, the localization effects, induced by 
pinning, dominate over mean free path at $K=0$.

The dependence of $ZT$ on $\eta$ is  also shown in Fig.~\ref{fig6}.
We see that a decrease of $\eta$ generates a slow
growth of $ZT$ even if at so low value as $\eta=0.01$ we still
have $ZT<2$. Here, we give numerical values of $\eta$
in our computational units. It is more physical to look
of a dimensionless ratio $\eta/\omega_0$ where $\omega_0$
is a maximal frequency of small oscillations near 
a vicinity of the Aubry ground state at $K = K_c$.
According to the results \cite{fki} we have
$\omega_0 \approx 2 \sqrt{K_c} \approx 0.4 $.
Thus all our data are obtained in the regime of 
long relaxation time scale ($\eta/\omega_0 \ll 1$). 
Also data obtained for longer chains 
$N=89$, $M=55$ and $N=144$, $M=89$ give 
no significant variation of $ZT$ with 
chain length (see Figs.~\ref{fig5},\ref{fig6}).

\begin{figure}[h]
\begin{center}
\includegraphics*[width=8.0cm]{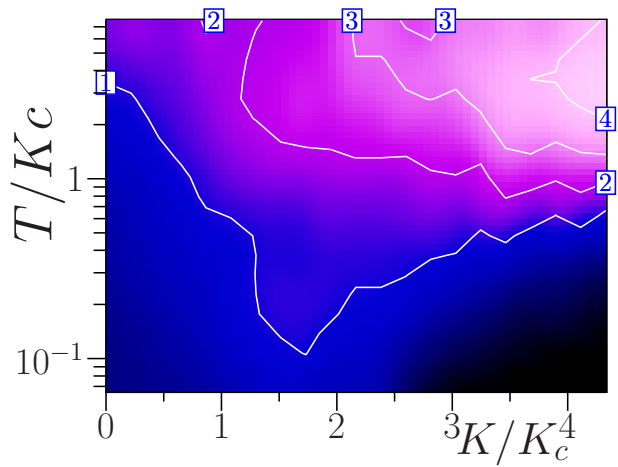}
\end{center}
\vglue -0.3cm
\caption{
Dependence of $ZT$ on $K/K_c$ and $T/K_c$
shown by color changing from $ZT=0$
(black) to maximal $ZT=4.5$ (light rose);
contour curves show values $ZT=1, 2, 3, 4$.
Here $\eta=0.02$, $N=34$, $M=21$. } 
\label{fig7}
\end{figure}

The global dependence of $ZT$ on $K/K_c$ and $T/K_c$ 
is presented in Fig.~\ref{fig7}
for the investigated parameter range
$T/K_c < 9, K/K_c \leq 4.5$.
The maximal value $ZT \approx 4.5$ is reached at
largest
$K/K_c \approx 4.5$ and $T/K_c \approx 4$.
However, at such large values of $K, T$
we start to enter in the regime
of potential and temperature being larger than
the Coulomb energy $E_W =\nu_g/2\pi$ 
so that it may be difficult to find materials
which realize effectively such a strong potential.
For a more realistic condition $T\leq K_c$
we have $ZT<2$.
We also note that for $K> 5 K_c \approx E_W$
the electrons are located in 
a strongly pinned phase with very strong
fluctuations of
transitions between
different minima
at $T \leq K_c$.

\begin{figure}[h]
\begin{center}
\includegraphics*[width=4cm]{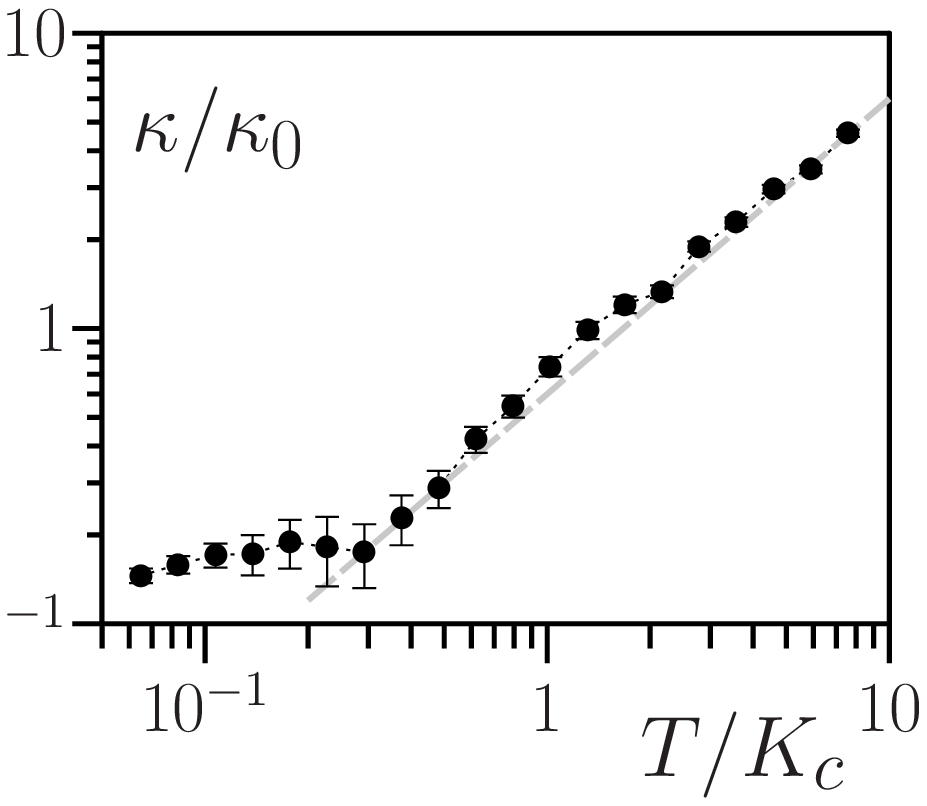}
\includegraphics*[width=4cm]{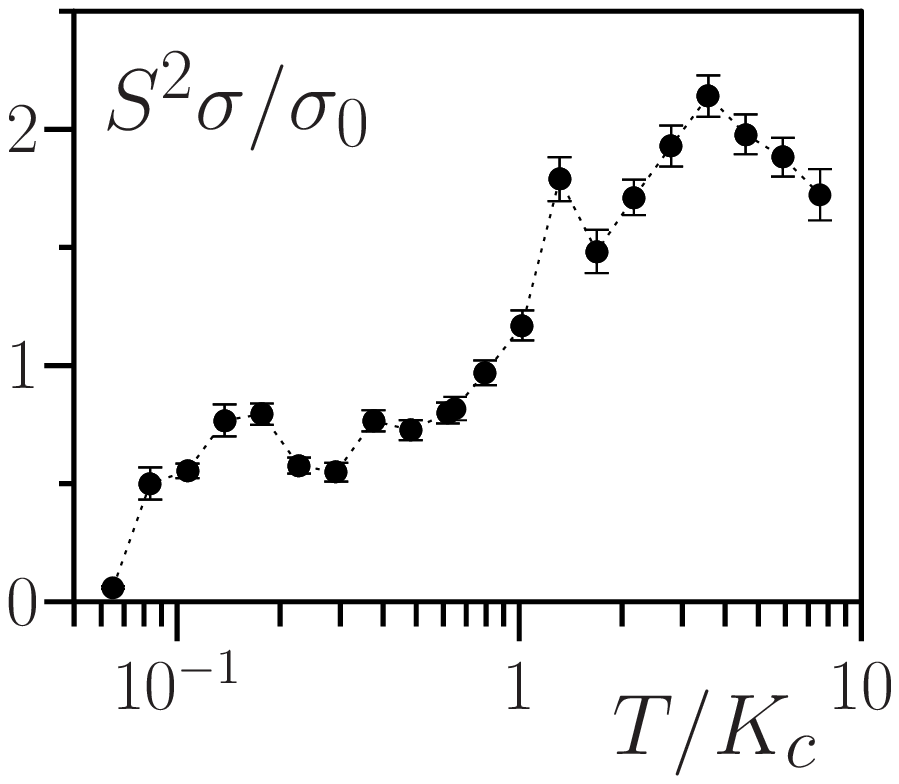}
\end{center}
\vglue -0.3cm
\caption{{\it Left panel}: Rescaled thermal conductivity
$\kappa/\kappa_0$ as a function of rescaler temperature  $T/K_c$,
to adapt an eye
the straight dashed line shows the dependence $\kappa/\kappa_0= 0.6 T/K_c$;
{\it right panel}: same as in left panel for
$S^2 \sigma/\sigma_0$. Data are obtained at
$K/K_c=2.6$, $\eta=0.02$, $N=34$, $M=21$,
$\sigma_0=\nu_g/(2\pi \eta)$, $\kappa_0=\sigma_0 K_c$.}
\label{fig8}
\end{figure}

\begin{figure}[h]
\begin{center}
\includegraphics*[width=4cm]{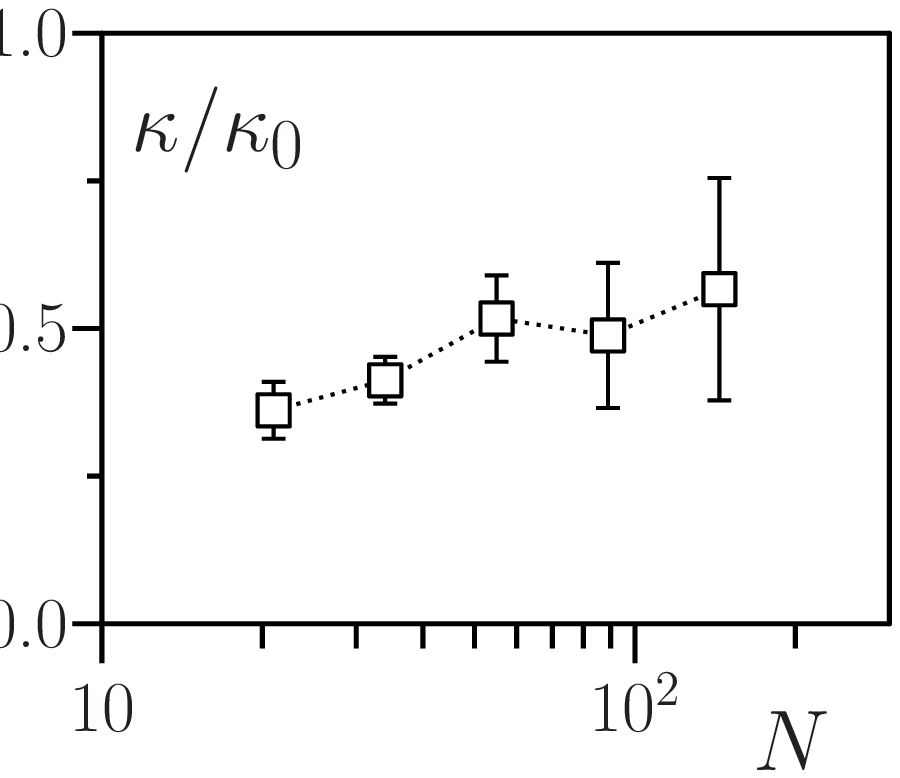}
\includegraphics*[width=4cm]{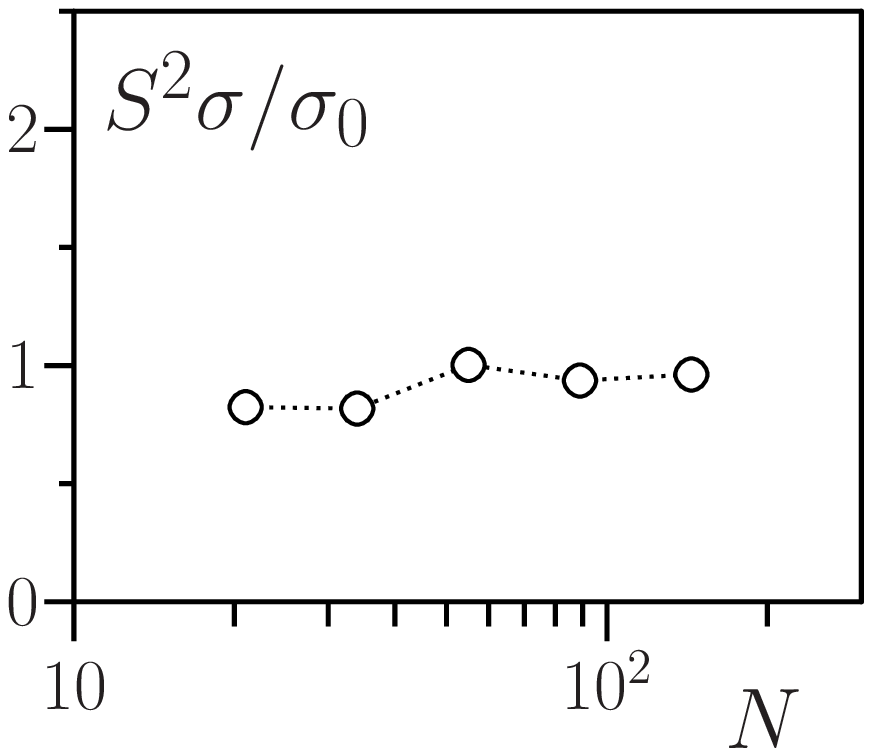}
\end{center}
\vglue -0.3cm
\caption{{\it Left panel}: Rescaled thermal conductivity
$\kappa/\kappa_0$ at different system sizes with number of electrons 
$N=21, 34, 55. 89, 144$;
{\it right panel}: same as in left panel for
$S^2 \sigma/\sigma_0$. Data are obtained at
$K/K_c=2.6$, $T/K_c=0.65$, $\eta=0.02$,
$\sigma_0=\nu_g/(2\pi \eta)$, $\kappa_0=\sigma_0 K_c$,
the system size $L=2\pi M$ is defined by the Fibonacci
value of $M$ at given $N$.}
\label{fig9}
\end{figure}

Additional data are presented in Fig.~\ref{fig8},
showing dependence of $\kappa$ and $S^2 \sigma$ 
on temperature $T$, and in Fig.~\ref{fig9},
showing independence of these quantities 
(within statistical errors)
of system size 
and number of electrons $N$.

\section{Discussion}

Our studies of the Wigner crystal
in a periodic potential show that
in the Aubry pinned phase at $K>K_c$
the Wigner
crystal has very larger Seebeck coefficients
$S$ which grow exponentially
with a decrease of temperature or
increase of the potential amplitude.
However, at the same time 
the charge and thermal conductivities
drop significantly. As a result,
for the all variety of cases studied we obtain the 
maximal value of $ZT  \leq 2$
for the realistic parameter range 
$K< 5 K_c, T< K_c$. Thus, there is
a rather nontrivial 
compensation of three quantities $S$, $\sigma$, $\kappa$
which determine the figure of merit, $ZT$.
In global, the pinned phase has larger $ZT$ 
values, compared to the sliding phase at $K<K_c$.
For high temperature $T \approx 4 K_c$
and  strongly pinned regime $K \approx 4K_c$
we obtain even $ZT \approx 4$.
However, it remains questionable if such high
potential amplitudes and temperatures are 
reachable in real materials.

We hope that it is possible to reach 
even larger $ZT$ in the Aubry pinned phase
at optimized system parameters.
We find that $ZT$ weakly increases with
a decrease of the seed
relaxation rate $\eta$. Thus a further
decrease of $\eta$ may allow to reach $ZT>3$
at low potential
amplitudes $K \approx 3 K_c$ and 
temperatures $T \approx K_c$.
However, special efforts should be performed to
determine this seed $\eta$ for real materials
since in the pinned phase the charge and thermal conductivities
drop significantly, compared to the sliding phase,
becoming practically independent of 
seed relaxation rate.

It is also possible that further temperature increase 
significantly above $T > 5 K_c$ may produce even $ZT>5$
at $K>5 K_c$.
However, the growth of $ZT$ with $T$ is slow,
being close to logarithmic growth,
so that such high $T$ and $K$ may be not interesting in
practice.

Thus the task to reach $ZT>3$ 
at low temperatures seems to be hard even in our 
simple model where the thermal conductivity of 
atomic lattice phonons is eliminated from the beginning
and only electronic conductivity contribution is left.
In this sense our model provides a superior bound for
$ZT$ factor in 1D. We expect that 
for the Wigner crystal in two- and three-dimensional
potentials the factor $ZT$  will be  
reduced, compared to 1D case,  since it will be 
more difficult to localize phonons of Wigner crystal.
Thus, in a certain sense we expect that
our model provides the most optimal conditions
for large $ZT$ values and still 
we remain at $ZT < 2$
for realistic not very high temperatures $T<K_c$. 

Finally we provide some physical 
values of our model parameters.
In physical units
we can estimate the critical potential amplitude
as $U_c =  K_c e^2/ (\epsilon d)$,
where $\epsilon$ is a dielectric constant,
 $\Delta x$ is a lattice period and 
$d=\nu \Delta x/2\pi$ is a rescaled lattice constant \cite{fki}. 
For values typical for a charge density wave  regime
\cite{brazovskii}
we have $\epsilon \sim 10$, $\nu \sim 1$, $\Delta x \sim 1 nm$
and  $U_c \sim 40 mV \sim 500 K$ 
so that the Aubry pinned phase
should be visible at room temperature.
The obtained $U_c$ value is rather high
that justifies the fact that we 
investigated thermoelectricity 
in the frame of classical mechanics
of interacting electrons.
In any case the real thermoelectric devices
should work at room temperature and in this
regime the classical treatment of 
electron transport can be considered as a good
first approximation.

We think that it would be useful to perform
experimental studies of electron transport in
a periodic potential. We hope that such type of experiments
can be possible with charge density waves
(see e.g. \cite{brazovskii} and Refs. therein),
strongly interacting electrons in 
ultraclean carbon nanotubes with  
interaction  energies of $100 mV$
\cite{nanotube}, 
experiments with electrons on a surface of liquid helium 
\cite{kono}, and cold ions in optical lattices \cite{haffner}. 

The research of OVZ was partially supported by the Ministry of
Education and Science of  Russian Federation.

\end{document}